\documentclass[11pt]{article}
\usepackage{latexsym,cite,amssymb,amsmath,times}
\usepackage{enumitem}
\usepackage{float}
\makeatletter
\@addtoreset{equation}{section} \makeatother

\input amssym.def
\input amssym.tex

\newcommand{\half}{{{\textstyle\frac{1}{2}}}}
\newcommand{\quarter}{{{\textstyle\frac{1}{4}}}}
\newcommand{\be}{\begin{equation}}
\newcommand{\ee}{\end{equation} }
\newcommand{\beqa}{\begin{eqnarray} }
\newcommand{\eeqa}{\end{eqnarray} }
\newcommand{\ba}{\begin{array}}
\newcommand{\ea}{\end{array}}

\newcommand{\SO}{\mathbf{SO}}
\newcommand{\Spin}{\mathbf{Spin}}

\newcommand{\mba}{{\mathbf{a}}}
\newcommand{\mbb}{{\mathbf{b}}}
\newcommand{\mbc}{{\mathbf{c}}}
\newcommand{\mbd}{{\mathbf{d}}}

\newcommand{\ODD}{\mathbf{O}(D,D)}

\newcommand{\etaodd}{{\cJ}}
\newcommand{\SOD}{{\SO(1,D{-1})}}
\newcommand{\oSOD}{{{\SO}(D{-1},1)}}
\newcommand{\SpinD}{{\Spin(1,D{-1})}}
\newcommand{\oSpinD}{{{\Spin}(D{-1},1)}}

\newcommand\cD{{\cal D}}

\newcommand\cH{{\cal H}}

\newcommand\cJ{{\cal J}}

\newcommand\cM{{\cal M}}

\newcommand\cP{{\cal P}}

\newcommand\bcP{{\bar{\cP}}}

\newcommand\hcL{{\hat{\cal L}}}

\newcommand\gammap{{\gamma^{\prime}{}}}
\newcommand\Dp{D^{\prime}}

\newcommand\cDp{\cD^{\prime}}

\newcommand\Vp{{V^{\prime}}{}}
\newcommand\brVp{{\brV^{\prime}}{}}
\newcommand\rhop{{\rho^{\prime}}{}}
\newcommand\psip{{\psi^{\prime}}{}}
\newcommand\brrhop{{\brrho^{\prime}}{}}
\newcommand\brpsip{{\brpsi^{\prime}}{}}
\newcommand\Phip{{\Phi^{\prime}}{}}
\newcommand\brPhip{{\brPhi^{\prime}}{}}
\def\brgammap{{\brgamma^{\prime}}{}}

\newcommand\dis{\displaystyle}

\def\tx{\tilde{x}}

\def\bre{\bar{e}}

\def\breta{\bar{\eta}}
\def\bralpha{\bar{\alpha}}
\def\brbeta{\bar{\beta}}
\def\brgamma{\bar{\gamma}}
\def\brdelta{\bar{\delta}}
\def\brrho{\bar{\rho}}
\def\brpsi{\bar{\psi}}

\def\brm{{\bar{m}}}

\def\brp{{\bar{p}}}
\def\brq{{\bar{q}}}
\def\brr{{\bar{r}}}
\def\brs{{\bar{s}}}
\def\bromega{{\bar{\omega}}}

\def\brPhi{{{\bar{\Phi}}}}

\def\brC{\bar{C}}
\def\brF{\bar{F}}

\def\brL{\bar{L}}

\def\brV{{\bar{V}}}
\def\brP{{\bar{P}}}

\def\brFp{{\brF^{\prime}{}}}
\def\Fp{{F^{\prime}{}}}
\def\Tw{{T}}

\newcommand{\DO}{\mathbf{\nabla}}

\newcommand{\na}{{\nabla}}
\newcommand{\trd}{{\bigtriangledown}}

\begin{document}
\begin{titlepage}
\title{%\vskip -60pt
%\vskip 20pt
\vskip 2cm
Incorporation of fermions  into double field theory \\~\\}
\author{\sc Imtak Jeon,${}^{\star}$\footnote{On leave of absence from Sogang University.}  \mbox{~~\,\,}
Kanghoon Lee${}^{\sharp}$ \mbox{~~\,\,}and\mbox{\,\,~~} Jeong-Hyuck Park${}^{\dagger}$}
\date{}
\maketitle \vspace{-1.0cm}
\begin{center}
~~~\\
${}^{\star}$CERN, Theory Division CH-1211 Geneva 23, Switzerland\\
~{}\\
${}^{\sharp}$Center for Quantum Spacetime, Sogang University,  Seoul 121-742, Korea\\
~\\
${}^{\dagger}$Department of Physics, Sogang University,  Seoul 121-742, Korea\\
%\texttt{imtak@sogang.ac.kr    ~~~~~park@sogang.ac.kr}
~{}\\
%\texttt{kanghoon@sogang.ac.kr}
%\texttt{}\\
%{\small{{Electronic correspondence:} \texttt{{{{~imtak,kanghoon,park@sogang.ac.kr}}}}}}\\
{{\texttt{~{{{imtak@sogang.ac.kr\,,~~kanghoon@sogang.ac.kr\,,~~park@sogang.ac.kr}}}}}}
~~~\\~\\~\\
\end{center}
\begin{abstract}
\vskip0.2cm
\noindent
Based on the stringy differential geometry we proposed  earlier, we incorporate  fermions such as gravitino and dilatino  into double field theory in a manifestly covariant manner with regard  to  $\ODD$ T-duality,  diffeomorphism, one-form gauge symmetry for $B$-field and a pair of local Lorentz symmetries.  We note    that  there are two kinds   of fermions in double field theory: $\ODD$  singlet   and non-singlet  which may be   identified,  respectively  as the common and the non-common fermionic sectors in type IIA and IIB supergravities.   For each kind,  we construct corresponding   covariant Dirac operators. Further, we derive  a simple  criterion for an $\ODD$ rotation to flip the chirality of the $\ODD$ non-singlet chiral  fermions, which  implies   the exchange of type IIA and IIB supergravities.  %%The criterion  depends on both the $\ODD$ group element  and the background fields. 
\end{abstract}

%%%
{\small
\begin{flushleft}
~~
\end{flushleft}}
\thispagestyle{empty}
%%%%
%%%
%%02.40.-k 	Geometry, differential geometry, and topology
%%11.25.-w 	Strings and branes
%%04.60.Cf 	Gravitational aspects of string theory
%%04.65.+e 	Supergravity 
%%%
\end{titlepage}
\newpage
\tableofcontents %%
%%\begin{document} --> JHEP
%\twocolumn
%%%%%%%%%%%%%%%%%%%%%%%%%%%%%%%%%%%%%%%%%%%%%%%%%%%%%%%%%%%%%%%%%%%%%%%%%%%%%%%%%%%%%%%%%%%%%%%%
%%%%%%%%%%%%%%%%%%%%%%%%%%%%%%%%%%%%%%%%%%%%%%%%%%%%%%%%%%%%%%%%%%%%%%%%%%%%%%%%%%%%%%%%%%%%%%%%
\section{Introduction} 
String theory possesses   T-duality and     imposes  $\ODD$ structure  on its $D$-dimensional  low energy effective actions~\cite{Buscher:1985kb,Buscher:1987sk,Buscher:1987qj,Giveon:1988tt}.  The $\ODD$ T-duality   can be conveniently described  if we  formally double  the spacetime dimension,   from  $D$ to $2D$,  with coordinates,  $x^{\mu}\rightarrow y^{A}=(\tx_{\mu},x^{\nu})$.  The new coordinates, $\tx_{\mu}$,  may  be viewed  as the canonical conjugates of the winding modes of closed strings, as noted   by Tseytlin and Siegel  in the early $90$'s~\cite{Tseytlin:1990nb,Tseytlin:1990va,Siegel:1993xq,Siegel:1993th}.     Recent developments  initiated by Hull and Zwiebach developed  this idea further,  in the name of Double Field Theory (DFT), by writing the $D$-dimensional effective action entirely in terms of the $2D$-dimensional language, \textit{i.e.~}$2D$ tensors~\cite{Hull:2009mi,Hull:2009zb,Hohm:2010jy,Hohm:2010pp} (see also \cite{Kwak:2010ew,Jeon:2010rw,Hohm:2010xe,Jeon:2011kp,Hohm:2011ex,Jeon:2011cn,Copland:2011yh,Thompson:2011uw,Hohm:2011dv,Albertsson:2011ux,Hohm:2011cp,ZwiebachLecture}).   Yet,  as a  field theory counterpart to  the level matching condition in closed string theories,   it is required that all the  fields   as well as  all of their possible products should be  annihilated by the $\ODD$ d'Alembert operator, $\partial^{2}=\partial_{A}\partial^{A}$,
\be
\ba{ll}
\partial^{2}\Phi\equiv 0\,,~~~~&~~~~
\partial_{A}\Phi_{1}\partial^{A}\Phi_{2}\equiv 0\,.
\ea
\label{constraint}
\ee
This `{level matching constraint}'  actually   means  that  the theory is not truly doubled:  there is a choice of coordinates $(\tx^{\prime},x^{\prime})$, related to the original coordinates $(\tx,x)$, by an  
$\ODD$ rotation, in which all the  fields do not depend on the $\tx^{\prime}$ coordinates~\cite{Hohm:2010jy}.  Henceforth, the equivalence symbol, `$\equiv$', means an equality up to the   constraint (\ref{constraint}), or simply up to the winding coordinate independency,  \textit{i.e.} $\frac{\partial~~}{\partial\tx_{\mu}}\equiv0$.  \newline

With the spacetime dimension formally doubled  in double field theory,   T-duality is   realized   by an $\ODD$ rotation which  acts on the $2D$-dimensional vector  indices of an $\ODD$ covariant tensor  in a standard manner, 
\be
\ba{ll}
\Tw_{A_{1}A_{2}\cdots A_{n}}~~\longrightarrow~~
M_{A_{1}}{}^{B_{1}}M_{A_{2}}{}^{B_{2}}\cdots M_{A_{n}}{}^{B_{n}}\Tw_{B_{1}B_{2}\cdots B_{n}}\,,
~~~&~~~M\in\ODD\,,
\ea
\ee
where the $\ODD$  group is defined by the invariance of a  constant metric, 
%%%
%%$\Big(\tiny\ba{cc}{\bf{0}}&{\bf{1}}\\{\bf{1}}&{\bf{0}}\ea\Big)$, 
%%%
\be
\ba{ll}
M_{A}{}^{C}M_{B}{}^{D}\etaodd_{CD}=\etaodd_{AB}\,,~~~~&~~~~\etaodd_{AB}:={\small{{{\left(\ba{cc}0&1\\1&0\ea\right)}}\,.}}
\ea
\label{ODDeta}
\ee
Without imposing the level matching constraint, the $\ODD$ transformation would naturally  correspond to a Noether symmetry of
the $2D$-dimensional field theory. However, with  the constraint, the double field theory is, by nature, $D$-dimensional   living  on a $D$-dimensional hyperplane. As the $\ODD$ transformation then rotates the entire hyperplane, the $\ODD$ rotation acts \textit{a priori} as a `duality' rather than a `Noether symmetry' of the $D$-dimensional theory. After further dimensional reductions, it becomes a Noether symmetry of the reduced action, as can be seen in \textit{e.g.}~\cite{Cremmer:1978km,Cremmer:1978ds,Buscher:1985kb,Buscher:1987sk,Buscher:1987qj,Maharana:1992my,Kleinschmidt:2004dy}.\\

\indent Further, in DFT the  $D$-dimensional diffeomorphism, $x^{\mu}\rightarrow x^{\mu}+\delta x^{\mu}$, and the one-form gauge symmetry of the two-form  gauge field, $B_{\mu\nu}\rightarrow B_{\mu\nu}+\partial_{\mu}\Lambda_{\nu}-\partial_{\nu}\Lambda_{\mu}$,  are  naturally combined  into what  we may  call `double-gauge symmetry'  (denoted by  `$\delta_{X}$'). By definition, the double-gauge transformation of a \textit{double-gauge covariant tensor}  is generated   by  the \textit{Dorfman derivative} or  \textit{generalized Lie derivative}, \textit{i.e.~}``$\delta_{X}=\hcL_{X}$",  whose  definition reads~\cite{Courant,Siegel:1993th,Gualtieri:2003dx,Grana:2008yw,Hohm:2010pp},
\be
\hcL_{X}\Tw_{A_{1}\cdots A_{n}}:=X^{B}\partial_{B}\Tw_{A_{1}\cdots A_{n}}+\omega_{{\scriptscriptstyle{T\,}}}\partial_{B}X^{B}\Tw_{A_{1}\cdots A_{n}}+\sum_{i=1}^{n}(\partial_{A_{i}}X_{B}-\partial_{B}X_{A_{i}})\Tw_{A_{1}\cdots A_{i-1}}{}^{B}{}_{A_{i+1}\cdots  A_{n}}\,.
\label{tcL}
\ee
Here $\omega_{{\scriptscriptstyle{T\,}}}$ is the given weight of an $\ODD$ covariant tensor, $\Tw_{A_{1}\cdots A_{n}}$, and  $X^{A}$ is the double-gauge symmetry  parameter whose  half  components are for  the  one-form gauge symmetry and the other half are  for the diffeomorphism,
\be
X^{A}=\left(\Lambda_{\mu}\,,\delta x^{\nu}\right)\,.
\label{XLdx}
\ee
As the generalized Lie derivative differs from the ordinary Lie derivative,  the underlying differential geometry of DFT should be beyond Riemann~\cite{Courant,Hitchin:2004ut,Hitchin:2010qz,Gualtieri:2003dx,Grana:2008yw,Jeon:2010rw,Jeon:2011cn,Coimbra:2011nw}.  Generally speaking,    while the fundamental object in Riemannian geometry is a metric, closed string theories call for
us to put the $B$-field and a scalar dilaton on an equal footing with the metric,  and  hence call for new geometry. \newline

In our previous works~\cite{Jeon:2010rw,Jeon:2011cn}, we
proposed a novel differential geometry for  double field theory that treats the three objects in a unified manner and manifests  
$\ODD$ T-duality, the double-gauge symmetry,    and also  a pair of local Lorentz symmetries simultaneously.  The key concept  therein  is  `semi-covariant derivative' that we review later.
\begin{table}[H]
\begin{center}
{\it{\small{\begin{itemize}
\item $\ODD$ T-duality
\item Gauge symmetries 
\begin{enumerate}
\item Double-gauge symmetry
\begin{itemize}
\item Diffeomorphism
\item One-form gauge symmetry
\end{itemize}
\item A pair of Local    Lorentz  symmetries, ${\SOD\times\oSOD}$
\end{enumerate}
\end{itemize}}}}
\caption{T-duality and gauge symmetries in DFT.  }
\label{TABsymmetry}
\end{center}
\end{table}

\indent In this paper,  utilizing  the semi-covariant derivative approach,  we   incorporate  fermions, such as gravitino and dilatino,  into double field theory.    Especially we construct     covariant  DFT Dirac operators  that are manifestly  compatible with all the symmetries in  Table \ref{TABsymmetry}.   Upon the level matching constraint (\ref{constraint}) and in terms of the undoubled $D$-dimensional component fields, our Dirac operators reduce to those found recently by Coimbra, Strickland-Constable and  Waldram as for  the unifying  reformulation of type IIA and IIB supergravities~\cite{Coimbra:2011nw}.\\

Further we show  that  there are two kinds  of fermions in double field theory: 
\begin{enumerate}
\item  $\ODD$ singlet fermions that, in our notation,   consist of  `unprimed'  gravitino and dilatino. Their   local Lorentz indices (spinorial and vectorial)   are singlet  under $\ODD$ T-duality.  They couple to 	`unprimed double-vielbein'~\cite{Jeon:2011cn}, 
\be
\ba{lll}
\left(\,\psi_{\brp}^{\,\alpha}, \,\rho^{\alpha}\,\right)~~~~&\Longleftrightarrow&~~~~\left(\,V_{Ap},\,\brV_{B\brp}\,\right)\,.
\ea
\ee
The common  fermionic sector of  type IIA and IIB supergravities may be identified as  our unprimed fermions. 

\item  $\ODD$ non-singlet fermions that consist of    `primed' gravitino and dilatino. Their   local Lorentz indices transform nontrivially under $\ODD$ T-duality. They couple to  `primed double-vielbein', 
\be
\ba{lll}
\left(\,\psip_{\brp}^{\,\alpha}, \,\rhop^{\alpha}\,\right)~~~~&\Longleftrightarrow&~~~~\left(\,\Vp_{A\brp},\,\brVp_{Bp}\,\right)\,.
\ea
\label{primedpsiV}
\ee
The non-common   fermions of the opposite chiralities in    type IIA and IIB supergravities correspond to our primed fermions. 
\end{enumerate}

We also present   a criterion for  $\ODD$ rotations to flip the chirality of the primed  fermions, which turns out to depend on both the $\ODD$ group element  and the background fields.  This  generalizes,  in a unifying manner,  the earlier works      by Hassan in 90's~\cite{Hassan:1994mq,Hassan:1999bv,Hassan:1999mm}.\\

The organization of the present paper is as follows. To start, in section \ref{SECconv} we set up  our conventions including   the indices  used  for each  representation of the symmetries  in Table \ref{TABsymmetry}.  In section \ref{SECtwotype}, 
after reviewing    the two types of the double-vielbeins  from  \cite{Jeon:2011cn}, we analyze   their finite $\ODD$ transformations.  In section \ref{SECDirac}, utilizing  the semi-covariant derivative,  we construct the covariant Dirac operators for each type of the fermions and derive the   criterion for the primed  fermions to flip  their  chiralities under $\ODD$ T-duality.  
Section \ref{SECconclusion} contains the summary and comments.   \\~\\

%%%%%%%%%%%%%%%%%%%%%%%%%%%%%%%%%%%%%%%%%%%%%%%%%%%%%%%%%%%%%%%%%%%%%%%%%%%%%%%%%%%%%%%%%%%%%%%%%%%%%%%%%%%%%%%%%%%%%%%%%%%%%%%%%%%%%%%%%%%%%%%%%%%%%%%%%%%%%%%%%%%%%%%%%%%%%%%%%%%%%%%%%%%%%%%%
\section{Conventions\label{SECconv}} 
In Table \ref{TABindices}, we  summarize   our conventions for  indices and   metrics used  for each  representation of the symmetries listed   in Table \ref{TABsymmetry}.\footnote{Note the opposite signatures chosen  for  $\eta$ and $\breta$, \textit{i.e.~}mostly plus \textit{vs.~}mostly minus (\textit{cf.~}\cite{Jeon:2011cn}).}
\begin{table}[H]
\begin{center}
\begin{tabular}{||c|c|c||}
\hline
~~~indices~~~&~~representation~~&~~~metric~~~\\
\hline
$A,B,\cdots$~&~double-gauge vector~&$\cJ_{AB}$ in Eq.(\ref{ODDeta})\\
$p,q,\cdots$~&~$\SOD$  vector~&$\eta_{pq}=\mbox{diag}(-++\cdots+)$ \\
$\alpha,\beta,\cdots$~&~$\SpinD$  spinor~&
$C_{\alpha\beta}$  in Eq.(\ref{CcM})\\
$\brp,\brq,\cdots$~&~$\oSOD$  vector~&$\breta_{\brp\brq}=\mbox{diag}(+--\cdots-)$ \\
$\bralpha,\brbeta,\cdots$~&~$\oSpinD$  spinor~&
$\brC_{\bralpha\brbeta}$  in Eq.(\ref{CcM})\\
\hline
\end{tabular}
\caption{Indices used for each symmetry representation and the relevant    metrics that raise or lower the positions of them.  While $\ODD$  acts always on the double-gauge vector  indices (capital Roman), it may also rotate other indices of the primed fields  (\ref{primedpsiV}). It is  the characteristic feature of DFT that, although the $\ODD$ metric $\cJ_{AB}$ (\ref{ODDeta}) is a  constant `flat' one, the corresponding `flat' indices, $A,B,\cdots,$ decompose into $D$-dimensional curved spacetime vector and one-form indices, as in (\ref{XLdx}), (\ref{Vform}), (\ref{Vpform}), \textit{etc.}~\cite{Hull:2009mi,Hull:2009zb,Hohm:2010jy,Hohm:2010pp}.}
\label{TABindices}
\end{center}
\end{table}
For the application of our formalism to type IIA and IIB supergravities, in this paper we focus on `even' $D$-dimensional Minkowskian  spacetime   that admits  Majorana-Weyl spinors, \textit{i.e.~}$D\approx 2$ mod $8$.  \\

For the two Minkowskian  metrics, $\eta_{pq}$ and $\breta_{\brp\brq}$,   we introduce  separately the corresponding `real' gamma matrices: $(\gamma^{p})^{\alpha}{}_{\beta}$ and $(\brgamma^{\brp})^{\bralpha}{}_{\brbeta}$ satisfying 
\be
\ba{ll}
\gamma^{p}=(\gamma^{p})^{\ast}\,,~~~~&~~~~
\gamma^{p}\gamma^{q}+\gamma^{q}\gamma^{p}=2\eta^{pq}\,,\\
\brgamma^{\brp}=(\brgamma^{\brp})^{\ast}\,,~~~~&~~~~
\brgamma^{\brp}\brgamma^{\brq}+\brgamma^{\brq}\brgamma^{\brp}=2\breta^{\brp\brq}\,.
\ea
\label{gammabr}
\ee  
Their  charge conjugation matrices,  $C_{\alpha\beta}$ and $\brC_{\bralpha\brbeta}$, meet\footnote{A possible relation between the unbarred and barred real gamma matrices is   to identify  $\brgamma^{\brp}$ with $\gamma^{p}\gamma^{(D+1)}$, and $\brC$  with $C$. However, we do not need to impose this identification   in the present paper. } 
\be
\ba{l}
(C\gamma^{p_{1}p_{2}\cdots p_{n}})_{\alpha\beta}=-(-1)^{n(n+1)/2}(C\gamma^{p_{1}p_{2}\cdots p_{n}})_{\beta\alpha}\,,\\
(\brC\brgamma^{\brp_{1}\brp_{2}\cdots\brp_{n}})_{\bralpha\brbeta}=-(-1)^{n(n+1)/2}(\brC\brgamma^{\brp_{1}\brp_{2}\cdots\brp_{n}})_{\brbeta\bralpha}\,,
\ea
\label{CcM}
\ee
and define the charge-conjugated  spinors. For the  unprimed and primed   $\SpinD$ spinors we have
\be
\ba{llll}
\brpsi_{\brp\alpha}=\psi_{\brp}^{\,\beta}C_{\beta\alpha}\,,~~~~&~~~~
\brrho_{\alpha}=\rho^{\beta} C_{\beta\alpha}\,,~~~~&~~~~
\brpsip_{\brp\alpha}=\psip_{\brp}^{\,\beta}C_{\beta\alpha}\,,~~~~&~~~~
\brrhop_{\alpha}=\rhop^{\beta}C_{\beta\alpha}\,.
\ea
\ee
\\
We also set, in order to specify  the chirality of the Weyl spinors,
\be
\ba{ll}
\gamma^{(D+1)}:=\gamma^{012\cdots D-1}\,,~~~~&~~~~\brgamma^{(D+1)}:=\brgamma^{012\cdots D-1}\,,
\ea
\label{g5}
\ee
that satisfy 
\be
\ba{llll}
\gamma^{p}\gamma^{(D+1)}+\gamma^{(D+1)}\gamma^{p}=0\,,~&~~\left(\gamma^{(D+1)}\right)^{2}=1\,,~&~~\brgamma^{\brp}\brgamma^{(D+1)}+\brgamma^{(D+1)}\brgamma^{\brp}=0\,,~&~~\left(\brgamma^{(D+1)}\right)^{2}=1\,.
\ea
\ee
The unprimed fermions, $(\psi_{\brp}^{\,\alpha},\rho^{\alpha})$, are  set to be  Majorana-Weyl spinors of the  fixed chiralities, 
\be
\ba{ll}
\gamma^{(D+1)}\psi_{\brp}=+\psi_{\brp}\,,~~~~&~~~~\gamma^{(D+1)}\rho=-\rho\,.
\ea
\ee
On the other hand, the primed fermions, $(\psip_{\brp}^{\,\alpha},\rhop^{\alpha})$, are Majorana-Weyl spinors   possessing    either    the same chirality (as for   type IIB supergravity),
\be
\ba{ll}
\gamma^{(D+1)}\psip_{\brp}=+\psip_{\brp}\,,~~~~&~~~~\gamma^{(D+1)}\rhop=-\rhop\,,
\ea
\ee
or  the opposite chirality  (as for   type IIA supergravity), 
\be
\ba{ll}
\gamma^{(D+1)}\psip_{\brp}=-\psip_{\brp}\,,~~~~&~~~~\gamma^{(D+1)}\rhop=+\rhop\,.
\ea
\ee
The  chiralities of the primed fermions   may be flipped under  $\ODD$ T-duality,  as we shall see later. \\
\newpage

%%%%%%%%%%%%%%%%%%%%%%%%%%%%%%%%%%%%%%%%%%%%%%%%%%%%%%%%%%%%%%%%%%%%%%%%%%%%%%%%%%%%%%%%%%%%%%%%%%%%%%%%%%%%%%%%%%%%%%%%%%%%%%%%%%%%%%%%%%%%%%%%%%%%%%%%%%%%%%%%%%%%%%%%%%%%%%%%%%%%%%%%%%%%%%%%
\section{Two types of double-vielbeins  and  their $\ODD$ transformations \label{SECtwotype}}
\subsection{Primed and unprimed double-vielbeins}
There are two types of vielbeins in DFT~\cite{Jeon:2011cn}. We distinguish them here as unprimed double-vielbein,  $(V_{Ap},\brV_{B\brq})$,  and primed double-vielbein,\footnote{In \cite{Jeon:2011cn}, the latter was   called ``twin double-vielbein". }   $(\Vp_{A\brp}\,\brVp_{Bq})$.    They carry   opposite   local Lorentz vector indices. \\

In terms of the flat metrics in Table \ref{TABindices}, the  unprimed double-vielbein  satisfies   the following defining properties~\cite{Jeon:2011cn}:
\be
\ba{llll}
V_{Ap}V^{A}{}_{q}=\eta_{pq}\,,~~&~~V_{Ap}\brV^{A}{}_{\brq}=0\,,~~&~~
\brV_{A\brp}\brV^{A}{}_{\brq}=\breta_{\brp\brq}\,,~~&~~V_{Ap}V_{B}{}^{p}+\brV_{A\brp}\brV_{B}{}^{\brp}=\cJ_{AB}\,.
\ea
\label{defV}
\ee
Hence the double-vielbein   forms  a pair of  rank-two  projections~\cite{Jeon:2010rw},
\be
\ba{ll}
P_{AB}:=V_{A}{}^{p}V_{Bp}\,,~~~~&~~~~\brP_{AB}:=\brV_{A}{}^{\brp}\brV_{B\brp}\,,
\ea
\ee
that are symmetric, orthogonal and complementary to each other,  
\be
\ba{lll}
P_{AB}=P_{BA}\,,~~&~~\brP_{AB}=\brP_{BA}\,,~~&~~P_{A}{}^{B}\brP_{B}{}^{C}=0\,,\\
P_{A}{}^{B}P_{B}{}^{C}=P_{A}{}^{C}\,,~~&~~\brP_{A}{}^{B}\brP_{B}{}^{C}=\brP_{A}{}^{C}\,,~~&~~P_{A}{}^{B}+\brP_{A}{}^{B}=\delta_{A}{}^{B}\,,
\ea
\label{symP2}
\ee
and further meet
\be
\ba{llll}
P_{A}{}^{B}V_{Bp}=V_{Ap}\,,~~&~~\brP_{A}{}^{B}\brV_{B\brp}=\brV_{A\brp}\,,~~&~~\brP_{A}{}^{B}V_{Bp}=0\,,~~&~~P_{A}{}^{B}\brV_{B\brp}=0\,.
\ea
\ee
The defining properties of the double-vielbein~(\ref{defV})   actually means then that, as a $2D\times 2D$ matrix, $(V_{A}{}^{p},\brV_{B}{}^{\brq})$ diagonalizes both the projectors $P_{AB}$ and $\brP_{AB}$, or equivalently both the $\ODD$ metric $\cJ_{AB}$ and the ``generalized metric" $\cH_{AB}:=(P-\brP)_{AB}$, as follows~\cite{Jeon:2011cn},
\be
\ba{ll}
\cJ={\Big(V\,,\brV\Big)}\left(\ba{cc}\eta&0\\0&\breta\ea\right){\Big(V\,,\brV\Big)}^{t}\,,~~~&~~~
\cH={\Big(V\,,\brV\Big)}\left(\ba{cc}\eta&\,0\\0&-\breta\ea\right){\Big(V\,,\brV\Big)}^{t}\,.
\ea
\label{diagJH}
\ee
Assuming that the upper half blocks are non-degenerate,  the unprimed double-vielbein takes the following most general form~\cite{Jeon:2011cn},\footnote{It is worth while to note that  (up to the upper half block non-degeneracy assumption),  (\ref{Vform}) is the most general form of the double-vielbein parametrization  that diagonalizes both  the $\ODD$ metric, $\cJ_{AB}$, and the generalized metric, $\cH_{AB}$, as in (\ref{diagJH}).  For other parametrization  that diagonalizes  $\cH_{AB}$ only, see the early  work by Maharana and  Schwarz~\cite{Maharana:1992my}. }
\be
\ba{ll}
V_{Ap}=\textstyle{\frac{1}{\sqrt{2}}}{{\left(\ba{c} (e^{-1})_{p}{}^{\mu}\\(B+e)_{\nu p}\ea\right)}}\,,~~~~
&~~~~\brV_{A{\brp}}=\textstyle{\frac{1}{\sqrt{2}}}\left(\ba{c} (\bre^{-1})_{\brp}{}^{\mu}\\(B+\bre)_{\nu{\brp}}\ea\right)\,.
\ea
\label{Vform}
\ee
Here  $e_{\mu}{}^{p}$ and  $\bre_{\nu}{}^{{\brp}}$ are two copies of  the $D$-dimensional   vielbein  corresponding  to  the same  spacetime metric in the following manner,   
\be
e_{\mu}{}^{p}e_{\nu}{}^{q}\eta_{pq}=-\bre_{\mu}{}^{{\brp}}\bre_{\nu}{}^{\brq}\breta_{\brp\brq}=g_{\mu\nu}\,,
\ee
and $B_{\mu\nu}$ corresponds to the Kalb-Ramond two-form gauge field.  
We also set  in (\ref{Vform}),
\be
\ba{ll}
B_{\mu p}=B_{\mu\nu}(e^{-1})_{p}{}^{\nu}\,,~~~~&~~~~B_{\mu\brp}=B_{\mu\nu}(\bre^{-1})_{{\brp}}{}^{\nu}\,.
\ea
\ee
In particular, $(\bre^{-1}e)_{\brp}{}^{p}$ and $(e^{-1}\bre)_{p}{}^{\brp}$ are local Lorentz transformations, 
\be
\ba{ll}
(\bre^{-1}e)_{\brp}{}^{p}(\bre^{-1}e)_{\brq}{}^{q}\eta_{pq}=-\breta_{\brp\brq}\,,~~~~&~~~~
(e^{-1}\bre)_{p}{}^{\brp}(e^{-1}\bre)_{q}{}^{\brq}\breta_{\brp\brq}=-\eta_{pq}\,.
\ea
\label{ebreeta}
\ee

\indent Now, having the explicit form of the unprimed  double-vielbein (\ref{Vform}), we  are able to  define  the `primed  double-vielbein',
\be
\ba{ll}
\Vp_{A\brp}:=(\bre^{-1}e)_{\brp}{}^{p}V_{Ap}=\textstyle{\frac{1}{\sqrt{2}}}{{\left(\ba{c} (\bre^{-1})_{\brp}{}^{\mu}\\(B-\bre)_{\nu\brp}\ea\right)}}\,,~~~
&~~~\brVp_{Ap}:=(e^{-1}\bre)_{p}{}^{\brp}\brV_{A\brp}=\textstyle{\frac{1}{\sqrt{2}}}\left(\ba{c} (e^{-1})_{p}{}^{\mu}\\(B-e)_{\nu p}\ea\right)\,.
\ea
\label{Vpform}
\ee
They satisfy, parallel  to (\ref{defV}),
\be
\ba{llll}
\Vp_{A\brp}\Vp^{A}{}_{\brq}=-\breta_{\brp\brq}\,,~&~\Vp_{A\brp}\brVp^{A}{}_{q}=0\,,~&~
\brVp_{Ap}\brVp^{A}{}_{q}=-\eta_{pq}\,,~&~\Vp_{A\brp}\Vp_{B}{}^{\brp}+\brVp_{Ap}\brVp_{B}{}^{p}=-\cJ_{AB}\,,
\ea
\ee
and 
\be
\ba{cc}
-\Vp_{A}{}^{\brp}\Vp_{B\brp}=V_{A}{}^{p}V_{Bp}=P_{AB}\,,~~~~&~~~~-\brVp_{A}{}^{p}\brVp_{Bp}=\brV_{A}{}^{\brp}\brV_{B\brp}=\brP_{AB}\,,\\
P_{A}{}^{B}\Vp_{B\brp}=\Vp_{A\brp}\,,~~~~&~~~~\brP_{A}{}^{B}\brVp_{Bp}=\brVp_{Ap}\,.
\ea
\ee
~\\

%%%%%%%%%%%%%%%%%%%%%%%%%%%%%%%%%%%%%%%%%%%%%%%%%%%%%%%%%%%%%%%%%%%%%%%%%%%%%%%%%%%%%%%%%%%%%%%%%%%%%%%%%%%%%%%%%%%%%%%%%%%%%%%%%%%%%%%%%%%%%%%%%%%%%%%%%%%%%%%%%%%%%%%%%%%%%%
\subsection{$\ODD$  rotations of the double-vielbeins}
Both the primed and unprimed double-vielbeins are covariant with respect to  the local Lorentz symmetries and  the double-gauge symmetry \textit{i.e.}  ``$\delta_{X}\equiv\hcL_{X}$". What make  them distinguishable    are their $\ODD$ T-duality transformations.  Once we set the unprimed double-vielbein  to be a covariant $\ODD$ vector,
\be
\ba{ll}
V_{Ap}~~\longrightarrow~~M_{A}{}^{B}V_{Bp}\,,~~~~&~~~~\brV_{A\brp}~~\longrightarrow~~M_{A}{}^{B}\brV_{B\brp}\,,
\ea
\label{vecunp}
\ee
the primed double-vielbein cannot  transform as an $\ODD$ vector: Its Lorentz vector indices must be   rotated too, as first noted in \cite{Jeon:2011cn} for infinitesimal $\ODD$ transformations.    Below we analyze their `finite' $\ODD$ transformations, for later discussion   on the chirality change of the primed fermions under $\ODD$ T-duality. \\

If we explicitly parametrize         a generic  $\ODD$ element as
\be
M_{A}{}^{B}=\left(\ba{cc}\mba^{\mu}{}_{\nu}&\mbb^{\mu\sigma}\\ 
\mbc_{\rho\nu}&\mbd_{\rho}{}^{\sigma}\ea\right)\,,
\label{Mpr}
\ee
the defining property of the $\ODD$ group  (\ref{ODDeta}) implies
\be
\ba{llll}
\mba\mbb^{t}+\mbb\mba^{t}=0\,,~~~~&~~~~\mbc\mbd^{t}+\mbd\mbc^{t}=0\,,~~~~&~~~~\mba\mbd^{t}+\mbb\mbc^{t}=1\,.
\ea
\label{mbab}
\ee
From the vectorial $\ODD$ transformation rule (\ref{vecunp}) of the unprimed double-vielbein (\ref{Vform}), we note, among others,
\be
\ba{ll}
e^{-1}~~\longrightarrow~~e^{-1}\left[\mba^{t}+(g-B)\mbb^{t}\right]\,,~~~~&~~~~
\bre^{-1}~~\longrightarrow~~\bre^{-1}\left[\mba^{t}-(g+B)\mbb^{t}\right]\,,
\ea
\ee
and hence
\be
\ba{ll}
(e^{-1}\bre)_{p}{}^{\brp}~~\longrightarrow~~L_{p}{}^{q}(e^{-1}\bre)_{q}{}^{\brp}\,,~~~~&~~~~(\bre^{-1}e)_{\brp}{}^{p}~~\longrightarrow~~\brL_{\brp}{}^{\brq}(\bre^{-1}e)_{\brq}{}^{p}\,,
\ea
\label{eeLL}
\ee
where we set
\be
\ba{ll}
L=e^{-1}\left[\mba^{t}+(g-B)\mbb^{t}\right]\left[\mba^{t}-(g+B)\mbb^{t}\right]^{-1}e\,,
~~~~&~~~~
\brL=(\bre^{-1}e)L^{-1}(e^{-1}\bre)\,.
\ea
\label{LbrL}
\ee

The crucial properties of $L$ and $\brL$ are that they are local Lorentz transformations, 
\be
\ba{ll}
L_{p}{}^{r}L_{q}{}^{s}\eta_{rs}=\eta_{pq}\,,~~~~&~~~~\brL_{\brp}{}^{\brr}\brL_{\brq}{}^{\brs}\breta_{\brr\brs}=\breta_{\brp\brq}\,.
\ea
\label{LLLorentz}
\ee
These can be verified  directly   from (\ref{mbab}) and
\be
\left[\mba+\mbb(g+B)\right]g^{-1}\left[\mba^{t}+(g-B)\mbb^{t}\right]=
\left[\mba-\mbb(g-B)\right]g^{-1}\left[\mba^{t}-(g+B)\mbb^{t}\right]\,.
\ee
In fact,  from the consideration that  $(e^{-1}\bre)_{p}{}^{\brp}$ and $(\bre^{-1}e)_{\brp}{}^{p}$ themselves  are local  Lorentz transformations and also that this property must be preserved under $\ODD$ T-duality,  it   follows  naturally that  $L$ and $\brL$  must correspond to local Lorentz transformations.\\

Therefore, under $\ODD$ T-duality the primed double-vielbein transforms nontrivially as 
\be
\ba{ll}
\Vp_{A\brp}~~\longrightarrow~~M_{A}{}^{B}\brL_{\brp}{}^{\brq}\Vp_{B\brq}\,,~~~~&~~~~\brVp_{Ap}~~\longrightarrow~~M_{A}{}^{B}L_{p}{}^{q}\brVp_{Bq}\,,
\ea
\label{avecp}
\ee
where $L$ and $\brL$ are  local Lorentz transformations (\ref{LbrL})  depending  on both the $\ODD$ element, $M$, and the  backgrounds, $g_{\mu\nu}$, $B_{\mu\nu}$.\\

It is well known that even-dimensional irreducible gamma matrices are unique up to similarity transformations,  essentially   due to Schur's lemma. This implies,  for the cases of (\ref{gammabr}), (\ref{ebreeta}) and   (\ref{LLLorentz}), that   there must be similarity transformations, $S_{e}$ satisfying
\be
\ba{ll}
\brgamma^{\brp}(\bre^{-1}e)_{\brp}{}^{p}=S_{e}^{-1}(\gamma^{(D+1)}\gamma^{p})S_{e}\,,~~~~&~~~~\gamma^{(D+1)}\gamma^{p}(e^{-1}\bre)_{p}{}^{\brp}=S_{e}\brgamma^{\brp}S_{e}^{-1}\,,
\ea
\label{SgamS}
\ee
and  also  $S_{L}$, $S_{\brL}$ satisfying 
\be
\ba{ll}
\gamma^{q}L_{q}{}^{p}=S_{L}^{-1}\gamma^{p}S_{L}\,,~~~~&~~~~
\brgamma^{\brq}\brL_{\brq}{}^{\brp}=S_{\brL}^{-1}\brgamma^{\brp}S_{\brL}\,.
\ea
\label{gLS}
\ee
From (\ref{g5}), (\ref{gLS}), we obtain 
\be
\ba{ll}
\gamma^{(D+1)}S_{L}=\det(L)\,S_{L}\gamma^{(D+1)}\,,
~~~~&~~~~\brgamma^{(D+1)}S_{\brL}=\det(\brL)\,S_{\brL}\brgamma^{(D+1)}\,,
\ea
\label{brS5}
\ee
where  from (\ref{LbrL}),
\be
\dis{
\det(L)=\det(\brL)^{-1}=\frac{\det\left[\mba+\mbb(g+B)\right]}{\det\left[\mba-\mbb(g-B)\right]}\,,}
\label{detL}
\ee
of which the  value must be either $+1$ or $-1$, since $L$ and $\brL$ are  local Lorentz transformations. 
Thus, if $\det(\brL)=+1$, $S_{\brL}$ commutes with $\brgamma^{(D+1)}$. Otherwise  \textit{i.e.~}$\det(\brL)=-1$,  they anti-commute. As we shall see in the following  section \ref{SECDirac}, this gives the necessary and sufficient condition  for the primed  fermions to flip  their  chiralities under $\ODD$ T-duality. \\
\\

In fact, using (\ref{brS5}), one can show that $S_{L}$ and  $S_{\brL}$ are related by
\be
S_{\brL}=\left\{\ba{ll}
S_{e}^{-1}S_{L}^{-1}S_{e}\quad&\quad\mbox{for\,~}\det(L)=+1\\
S_{e}^{-1}\gamma^{(D+1)}S_{L}^{-1}S_{e}\quad&\quad\mbox{for\,~}\det(L)=-1\,.
\ea
\right.
\label{SLS}
\ee

\newpage

%%%%%%%%%%%%%%%%%%%%%%%%%%%%%%%%%%%%%%%%%%%%%%%%%%%%%%%%%%%%%%%%%%%%%%%%%%%%%%%%%%%%%%%%%%%%%%%%%%%%%%%%%%%%%%%%%%%%%%%%%%%%%%%%%%%%%%%%%%%%%%%%%%%%%%%%%%%%%%%%%%%%%%%%%%%%%%%%%%%%%%%%%%%%%%%%
\section{Covariant Dirac operators\label{SECDirac}} 
In this section, utilizing  the  semi-covariant derivative in  Refs.\cite{Jeon:2010rw,Jeon:2011cn},  we construct  covariant Dirac operators for unprimed and primed fermions separately, and discuss the  chirality change of the primed fermions  under $\ODD$ T-duality.\\ 

%%%%%%%%%%%%%%%%%%%%%%%%%%%%%%%%%%%%%%%%%%%%%%%%%%%%%%%%%%%%%%%%%%%%%%%%%%%%%%%%%%%%%%%%%%%%%%%%%%%%%%%%%%%%%%%%%%%%%%%%%%%%%%%%%%%%%%%%%%%%%%%%%%%%%%%%%%%%%%%%%%%%%%%%%%%%%%%%%%%%%%%%%%%%%%%%%%%%%%%%%%%%%%%%%%%%%%%%%%%%%%%%%%%%%%%%%%%%%%%%%%%%%%%%%%%%%%%%%%%%%%%%%%%%%%%%%%
\subsection{Semi-covariant derivative for double-gauge symmetry: review}
By definition~\cite{Jeon:2010rw,Jeon:2011cn},  the  semi-covariant derivative  acts on  a generic $\ODD$ tensor density with weight, $\omega_{{\scriptscriptstyle{T\,}}}$,  as
\be
\DO_{C}\Tw_{A_{1}A_{2}\cdots A_{n}}:=\partial_{C}\Tw_{A_{1}A_{2}\cdots A_{n}}-\omega_{{\scriptscriptstyle{T\,}}}\Gamma^{B}{}_{BC}\Tw_{A_{1}A_{2}\cdots A_{n}}
+\sum_{i=1}^{n}\,\Gamma_{CA_{i}}{}^{B}\Tw_{A_{1}\cdots A_{i-1}BA_{i+1}\cdots A_{n}}\,,
\label{semi-covD}
\ee
and it annihilates  the pair of rank-two  projections and the DFT-dilaton (and hence the NS-NS sector completely), 
\be
\ba{lll}
\na_{A}P_{BC}=0\,,~~~~&~~~~\na_{A}\brP_{BC}=0\,,~~~~&~~~~
\na_{A}d:=-\half e^{2d}\na_{A}(e^{-2d})=\partial_{A}d+\half\Gamma^{B}{}_{BA}=0\,.
\ea
\label{naPPd}
\ee
Note that the DFT-dilaton, $d$,  is  related to the string dilaton, $\phi$, through~\cite{Hohm:2010jy}
\be
e^{-2d}=\sqrt{-g}e^{-2\phi}\,,
\ee
and hence $e^{-2d}=\sqrt{-g}e^{-2\phi}$ is a scalar density with  weight one. In fact, this is the only quantity having a nontrivial weight in this paper.\\

  It follows  from (\ref{naPPd}) that, the $\ODD$ metric is also  `{flat}' with respect to the  semi-covariant derivative,
\be
\na_{A}\cJ_{BC}=0\,,
\ee
which implies
\be
\Gamma_{ABC}=-\Gamma_{ACB}\,.
\label{symG}
\ee
Further, requiring
\be
\Gamma_{ABC}+\Gamma_{BCA}+\Gamma_{CAB}=0\,,
\label{torsionless}
\ee
and
\be
\ba{ll}
\cP_{CAB}{}^{DEF}\Gamma_{DEF}=0\,,~~~~&~~~~
\bcP_{CAB}{}^{DEF}\Gamma_{DEF}=0\,,
\ea
\label{RANK6used}
\ee
the connection is uniquely fixed to be~\cite{Jeon:2011cn}
\be
\ba{ll}
\Gamma_{CAB}=&2\left(P\partial_{C}P\brP\right)_{[AB]}+2\left({{\brP}_{[A}{}^{D}{\brP}_{B]}{}^{E}}-{P_{[A}{}^{D}P_{B]}{}^{E}}\right)\partial_{D}P_{EC}\\
{}&-\textstyle{\frac{4}{D-1}}\left(\brP_{C[A}\brP_{B]}{}^{D}+P_{C[A}P_{B]}{}^{D}\right)\left(\partial_{D}d+(P\partial^{E}P\brP)_{[ED]}\right)\,.
\ea
\label{Gammao}
\ee
In (\ref{RANK6used}),  $\cP_{CAB}{}^{DEF}$ and $\bcP_{CAB}{}^{DEF}$ are    rank-six  projections,
\be
\ba{l}
\cP_{CAB}{}^{DEF}:=P_{C}{}^{D}P_{[A}{}^{[E}P_{B]}{}^{F]}+\textstyle{\frac{2}{D-1}}P_{C[A}P_{B]}{}^{[E}P^{F]D}\,,\\
\bcP_{CAB}{}^{DEF}:=\brP_{C}{}^{D}\brP_{[A}{}^{[E}\brP_{B]}{}^{F]}+\textstyle{\frac{2}{D-1}}\brP_{C[A}\brP_{B]}{}^{[E}\brP^{F]D}\,,
\ea
\label{P6}
\ee
that are symmetric and traceless,
\be
\ba{ll}
{\cP_{CABDEF}=\cP_{DEFCAB}=\cP_{C[AB]D[EF]}\,,}~~&~~{\bcP_{CABDEF}=\bcP_{DEFCAB}=\bcP_{C[AB]D[EF]}\,,} \\
{\cP_{CAB}{}^{DEF}\cP_{DEF}{}^{GHI}=\cP_{CAB}{}^{GHI}\,,}~~&~~{\bcP_{CAB}{}^{DEF}\bcP_{DEF}{}^{GHI}=\bcP_{CAB}{}^{GHI}\,,}\\
{\cP^{A}{}_{ABDEF}=0\,,~~~~P^{AB}\cP_{ABCDEF}=0\,,}~~&~~
{\bcP^{A}{}_{ABDEF}=0\,,~~~~\brP^{AB}\bcP_{ABCDEF}=0\,.}
\ea
\label{symP6}
\ee
The symmetric properties, (\ref{symG}) and (\ref{torsionless}), enable us to replace  the ordinary derivatives in the definition
of the generalized Lie derivative (\ref{tcL})  by our semi-covariant derivatives (\ref{semi-covD}), \textit{i.e.~}$\hcL_{X}^{\partial}\rightarrow\hcL_{X}^{\na}$.   The additional constraints (\ref{torsionless}) and (\ref{RANK6used}) are analogue to the torsionless condition in Riemannian geometry that   uniquely picks up the  the  Levi-Civita connection.  In fact, assuming  the skew-symmetric property, $\Gamma_{ABC}=-\Gamma_{ACB}$ only,  the difference between  $\hcL_{X}^{\partial}$ and $\hcL_{X}^{\na}$ is given by the totally anti-symmetric part of the connection, 
\be
\dis{\left(\hcL_{X}^{\na}-\hcL_{X}^{\partial}\right)\Tw_{A_{1}\cdots A_{n}}=\sum_{i=1}^{n}\left(\Gamma_{A_{i}BC}+\Gamma_{BCA_{i}}+\Gamma_{CA_{i}B}\right)X^{C}\Tw_{A_{1}\cdots A_{i-1}}{}^{B}{}_{A_{i+1}\cdots A_{n}}\,,}
\ee
such that this difference might be used for the definition of ``torsion"~\cite{Coimbra:2011nw}. However we emphasize  that, the  symmetric properties (\ref{symG}), (\ref{torsionless}) are not sufficient enough to fix the connection uniquely:  the projective  condition (\ref{RANK6used}) must be  also imposed.  \\

\indent Under the double-gauge  transformations,  the  connection and  the semi-covariant derivative transform as
\be
\ba{l}
(\delta_{X}{-\hcL_{X}})\Gamma_{CAB}\equiv 2\big[(\cP{+\bcP})_{CAB}{}^{FDE}-\delta_{C}^{~F}\delta_{A}^{~D}\delta_{B}^{~E}\big]\partial_{F}\partial_{[D}X_{E]}\,,\\
\dis{(\delta_{X}{-\hcL_{X}})\na_{C}T_{A_{1}\cdots A_{n}}\equiv
\sum_{i=1}^{n}2(\cP{+\bcP})_{CA_{i}}{}^{BFDE}
\partial_{F}\partial_{[D}X_{E]}T_{A_{1}\cdots A_{i-1} BA_{i+1}\cdots A_{n}}\,.}
\ea
\label{noncov}
\ee
Hence, they  are not double-gauge covariant. We say, a tensor is double-gauge covariant if and only if its double-gauge transformation agrees  with  the   generalized Lie derivative,  \textit{i.e.~}`$\delta_{X}=\hcL_{X}$'.  Nonetheless,    the characteristic feature  of the semi-covariant  derivative is that,   combined with the projections,  it can  generate  various     fully covariant    quantities, and hence the name `semi-covariant':
\be
\ba{c}
P_{C}{}^{D}{\brP}_{A_{1}}{}^{B_{1}}{\brP}_{A_{2}}{}^{B_{2}}\cdots{\brP}_{A_{n}}{}^{B_{n}}
\DO_{D}T_{B_{1}B_{2}\cdots B_{n}}\,,\\
{\brP}_{C}{}^{D}P_{A_{1}}{}^{B_{1}}P_{A_{2}}{}^{B_{2}}\cdots P_{A_{n}}{}^{B_{n}}
\DO_{D}T_{B_{1}B_{2}\cdots B_{n}}\,,\\
P^{AB}{\brP}_{C_{1}}{}^{D_{1}}{\brP}_{C_{2}}{}^{D_{2}}\cdots{\brP}_{C_{n}}{}^{D_{n}}\DO_{A}T_{BD_{1}D_{2}\cdots D_{n}}\,,\\
\brP^{AB}{P}_{C_{1}}{}^{D_{1}}{P}_{C_{2}}{}^{D_{2}}\cdots{P}_{C_{n}}{}^{D_{n}}\DO_{A}T_{BD_{1}D_{2}\cdots D_{n}}\,,\\
P^{AB}{\brP}_{C_{1}}{}^{D_{1}}{\brP}_{C_{2}}{}^{D_{2}}\cdots{\brP}_{C_{n}}{}^{D_{n}}
\DO_{A}\DO_{B}T_{D_{1}D_{2}\cdots D_{n}}\,,\\
{\brP}^{AB}P_{C_{1}}{}^{D_{1}}P_{C_{2}}{}^{D_{2}}\cdots P_{C_{n}}{}^{D_{n}}
\DO_{A}\DO_{B}T_{D_{1}D_{2}\cdots D_{n}}\,.
\ea
\label{covariant}
\ee
~\\
\indent With the usual  curvature,
\be
R_{CDAB}=\partial_{A}\Gamma_{BCD}-\partial_{B}\Gamma_{ACD}+\Gamma_{AC}{}^{E}\Gamma_{BED}-\Gamma_{BC}{}^{E}\Gamma_{AED}\,,
\ee 
that turns out to be double-gauge non-covariant,  if we set 
\be
S_{ABCD}:=\half\left(R_{ABCD}+R_{CDAB}-\Gamma^{E}{}_{AB}\Gamma_{ECD}\right)\,,
\ee
a double-gauge  covariant rank two-tensor  and a double-gauge  covariant   scalar follow
\be
\ba{ll}
P_{I}{}^{A}\brP_{J}{}^{B}S_{AB}\,,~~~~&~~~~P^{AB}S_{AB}\equiv-\brP^{AB}S_{AB}\,.
\ea
\label{dgcov}
\ee
Here we put
\be
S_{AB}{=S_{BA}}{:=S^{C}{}_{ACB}}\,,
\ee
that turns out to be  traceless,
\be
{S^{A}{}_{A}}\equiv 0\,.
\label{trless}
\ee
In particular, the  covariant scalar  reduces to  the bosonic  closed string effective action upon the level matching constraint~(\ref{constraint})~\cite{Jeon:2011cn}, 
\be
2P^{AB}S_{AB}\equiv R_{g}+4\Box\phi
-4\partial_{\mu}\phi\partial^{\mu}\phi-\textstyle{\frac{1}{12}}H_{\lambda\mu\nu}H^{\lambda\mu\nu}\,.
\label{NSaction3}
\ee
~\\

%%%%%%%%%%%%%%%%%%%%%%%%%%%%%%%%%%%%%%%%%%%%%%%%%%%%%%%%%%%%%%%%%%%%%%%%%%%%%%%%%%%%%%%%%%%%%%%%%%%%%%%%%%%%%%%%%%%%%%%%%%%%%%%%%%%%%%%%%%%%%%%%%%%%%%%%%%%%%%%%%%%%%%%%%%%%%%%%%%%%%%%%%%%%%%%%%%%%%%%%%%%%%%%%%%%%%%%%%%%%%%%%%%%%%%%%%%%%%%%%%%%%%%%%%%%%%%%%%%%%%%%%%%%%%%%%%%
\subsection{Unprimed Dirac operators: $\ODD$ singlet}
For the $\ODD$ singlet fermions, \textit{i.e.~}unprimed fermions, $(\psi_{\brp}^{\,\alpha},\rho^{\alpha})$,  we focus on the following  differential operator,
\be
\cD_{A}:=\partial_{A}+\Gamma_{A}+\Phi_{A}+\brPhi_{A}=\na_{A}+\Phi_{A}+\brPhi_{A}=D_{A}+\Gamma_{A}\,,
\label{cDA}
\ee
where  $D_{A}$, is a  local Lorenz covariant, yet double-gauge non-covariant,  derivative  having  the connections,  $\Phi_{A}$ and  $\brPhi_{A}$  for $\SOD$ and $\oSOD$ respectively,
\be
D_{A}:=\partial_{A}+\Phi_{A}+\brPhi_{A}\,.
\ee
We view $\cD_{A}$,  as our  `master' unprimed,  semi-covariant derivative unifying $\na_{A}$ and $D_{A}$. We require it to annihilate the unprimed double-vielbein and the DFT-dilaton,  
\be
\ba{l}
\cD_{A}V_{Bp}=\partial_{A}V_{Bp}+\Gamma_{AB}{}^{C}V_{Cp}+\Phi_{Ap}{}^{q}V_{Bq}=0\,,\\
\cD_{A}\brV_{B\brp}=\partial_{A}\brV_{B\brp}+\Gamma_{AB}{}^{C}\brV_{C\brp}+\brPhi_{A\brp}{}^{\brq}\brV_{B\brq}=0\,,\\
\cD_{A}d=\na_{A}d:=-\half e^{2d}\na_{A}(e^{-2d})=\partial_{A}d+\half\Gamma^{B}{}_{BA}=0\,,
\ea
\label{VVd}
\ee
 as well as  all the  constant  metrics and  the gamma matrices in Table \ref{TABindices},
 \be
 \ba{lllll}
\cD_{A}\cJ_{BC}=0\,,~~~&~~~~\cD_{A}\eta_{pq}=0\,,~~~&~~~~\cD_{A}\breta_{\brp\brq}=0\,,~~~&~~~~
\cD_{A}C_{\alpha\beta}=0\,,~~~&~~~~\cD_{A}\brC_{\bralpha\brbeta}=0\,,\\
\multicolumn{5}{c}{\cD_{A}(\gamma^{p})^{\alpha}{}_{\beta}=\Phi_{A}{}^{p}{}_{q}(\gamma^{q})^{\alpha}{}_{\beta}
+\Phi_{A}{}^{\alpha}{}_{\delta}(\gamma^{p})^{\delta}{}_{\beta}-(\gamma^{p})^{\alpha}{}_{\delta}
\Phi_{A}{}^{\delta}{}_{\beta}=0\,,}\\
\multicolumn{5}{c}{\cD_{A}(\brgamma^{\brp})^{\bralpha}{}_{\brbeta}=\brPhi_{A}{}^{\brp}{}_{\brq}(\brgamma^{\brq})^{\bralpha}{}_{\brbeta}
+\brPhi_{A}{}^{\bralpha}{}_{\brdelta}(\brgamma^{\brp})^{\brdelta}{}_{\brbeta}-(\brgamma^{\brp})^{\bralpha}{}_{\brdelta}
\brPhi_{A}{}^{\brdelta}{}_{\brbeta}=0\,.}
\ea
\label{JeeCC}
\ee
It follows that 
\be
\ba{ll}
\cD_{A}P_{BC}=\na_{A}P_{BC}=0\,,~~~~&~~~~\cD_{A}\brP_{BC}=\na_{A}\brP_{BC}=0\,,
\ea
\ee
and as usual,\footnote{Here, for simplicity, we omit the possibility of adding a central term to the spin connections, \textit{i.e.} 
\[\Phi_{A}{}^{\alpha}{}_{\beta}=\quarter\Phi_{Apq}(\gamma^{pq})^{\alpha}{}_{\beta}+c\times\delta^{\alpha}{}_{\beta}\,.\] }
\be
\ba{llll}
\Phi_{Apq}=-\Phi_{Aqp}\,,~~&~~\brPhi_{A\brp\brq}=-\brPhi_{A\brq\brp}\,,~~&~~
\Phi_{A}{}^{\alpha}{}_{\beta}=\quarter\Phi_{Apq}(\gamma^{pq})^{\alpha}{}_{\beta}\,,~~&~~
\brPhi_{A}{}^{\bralpha}{}_{\brbeta}=\quarter\brPhi_{A\brp\brq}(\brgamma^{\brp\brq})^{\bralpha}{}_{\brbeta}\,.
\ea
\ee
Specifically the spin  connections are determined,  from (\ref{VVd}) with (\ref{Gammao}), by
\be
\ba{ll}
\Phi_{Apq}=V^{B}{}_{p}\na_{A}V_{Bq}\,,~~~~&~~~~
\brPhi_{A\brp\brq}=\brV^{B}{}_{\brp}\na_{A}\brV_{B\brq}\,,
\ea
\ee
such that
\be
\Gamma_{ABC}=V_{B}{}^{p}D_{A}V_{Cp}+\brV_{B}{}^{\brp}D_{A}\brV_{C\brp}\,.
\ee
From the consideration of   ${[\cD_{A},\cD_{B}]V_{Cp}=0}\,$ and $\,{[\cD_{A},\cD_{B}]\brV_{C\brp}=0}$, we may  derive the relations between the  covariant scalar, $P^{AB}S_{AB}$ (\ref{dgcov}), and the field strengths of the local Lorentz  connections,
\be
\ba{l}
P^{AB}S_{AB}=F_{ABpq}V^{Ap}V^{Bq}- \half\Gamma_{ABC}\Gamma^{AB}{}_{D}P^{CD}\,,\\
\brP^{AB}S_{AB}=\brF_{AB\brp\brq}\brV^{A\brp}\brV^{B\brq}- \half\Gamma_{ABC}\Gamma^{AB}{}_{D}\brP^{CD}\,,
\ea
\label{SF}
\ee
where, in fact  $P^{AB}S_{AB}=-\brP^{AB}S_{AB}$ due to  (\ref{trless}), and the field strengths are as usual, 
\be
\ba{l}
F_{ABpq}=\partial_{A}\Phi_{Bpq}-\partial_{B}\Phi_{Apq}+\Phi_{Apr}\Phi_{B}{}^{r}{}_{q}-\Phi_{Bpr}\Phi_{A}{}^{r}{}_{q}\,,\\
\brF_{AB\brp\brq}=\partial_{A}\brPhi_{B\brp\brq}-\partial_{B}\brPhi_{A\brp\brq}+\brPhi_{A\brp\brr}\brPhi_{B}{}^{\brr}{}_{\brq}-\brPhi_{B\brp\brr}\brPhi_{A}{}^{\brr}{}_{\brq}\,.
\ea
\label{FPhi}
\ee
~\\
\indent Though $\Phi_{Apq}$ and $\brPhi_{A\brp\brq}$ are not double-gauge covariant from (\ref{noncov}),\footnote{For double-gauge covariant yet $\ODD$ non-covariant connections for the local Lorentz symmetries, see our earlier work~\cite{Jeon:2011cn}. In this paper,  instead   we focus on  the  double-gauge semi-covariant and $\ODD$ covariant connections for the local Lorentz symmetries. }
\[
\ba{l}
(\delta_{X}-\hcL_{X})\Phi_{Apq}\equiv2\cP_{ABC}{}^{DEF}\partial_{D}\partial_{[E}X_{F]}V^{B}{}_{p}V^{C}{}_{q}\,,\\
(\delta_{X}-\hcL_{X})\brPhi_{A\brp\brq}\equiv2\bcP_{ABC}{}^{DEF}\partial_{D}\partial_{[E}X_{F]}\brV^{B}{}_{\brp}\brV^{C}{}_{\brq}\,,
\ea
\] 
with  (\ref{P6}),  the followings are so, \textit{i.e.~}`\,$\delta_{X}\equiv\hcL_{X}$\,',
\be
\ba{llllll}
\brP_{A}{}^{B}\Phi_{Bpq}\,,~&~P_{A}{}^{B}\brPhi_{B\brp\brq}\,,~&~\Phi_{A[pq}V^{A}{}_{r]}\,,~&~
\brPhi_{A[\brp\brq}\brV^{A}{}_{\brr]}\,,~&~\Phi_{Apq}V^{Ap}\,,~&~\brPhi_{A\brp\brq}\brV^{A\brp}\,.
\ea
\label{covPhi}
\ee 
This generalizes our earlier results in \cite{Jeon:2011cn} where only the first two in (\ref{covPhi}) were identified.\\

After all, the fully  covariant unprimed Dirac operators, with respect to all the symmetries in Table \ref{TABsymmetry},  are as follows
\be
\ba{llll}
\gamma^{A}\cD_{A}\rho\,,~~~~&~~~~\gamma^{A}\cD_{A}\psi_{\brp}\,,~~~~&~~~~\brV^{A}{}_{\brp}\cD_{A}\rho\,,~~~~&~~~~\brV^{A\brp}\cD_{A}\psi_{\brp}=\cD_{A}\psi^{A}\,.
\ea
\label{unprimedDirac}
\ee
Here   we  set for simplicity,
\be
\ba{ll}
\psi_{A}:=\brV_{A}{}^{\brp}\psi_{\brp}\,,~~~~&~~~~\gamma^{A}:=V^{A}{}_{p}\gamma^{p}\,,
\ea
\ee
such  that 
\be
\ba{ll}
\brV^{A}{}_{\brp}\psi_{A}=\psi_{\brp}\,,~~~~&~~~~\left\{\gamma^{A},\gamma^{B}\right\}=2P^{AB}\,.
\ea
\ee 
Writing explicitly,
\be
\ba{l}
\cD_{A}\rho=D_{A}\rho=(\partial_{A}+\quarter\Phi_{Apq}\gamma^{pq})\rho\,,\\
\cD_{A}\psi_{\brp}=D_{A}\psi_{\brp}=(\partial_{A}+\quarter\Phi_{Apq}\gamma^{pq})\psi_{\brp}
+\brPhi_{A\brp}{}^{\brq}\psi_{\brq}\,,\\
\cD_{A}\psi_{B}=(\partial_{A}+\quarter\Phi_{Apq}\gamma^{pq})\psi_{B}+\Gamma_{AB}{}^{C}\psi_{C}\,.
\ea
\ee
As all  the associated  fields in  (\ref{unprimedDirac})  are unprimed (unprimed double-vielbein and unprimed fermions),  
all the unprimed Dirac operators  are  $\ODD$ singlets. \\

%%%%%%%%%%%%%%%%%%%%%%%%%%%%%%%%%%%%%%%%%%%%%%%%%%%%%%%%%%%%%%%%%%%%%%%%%%%%%%%%%%%%%%%%%%%%%%%%%%%%%%%%%%%%%%%%%%%%%%%%%%%%%%%%%%%%%%%%%%%%%%%%%%%%%%%%%%%%%%%%%%%%%%%%%%%%%%%%%%%%%%%%%%%%%%%%%%%%%%%%%%%%%%%%%%%%%%%%%%%%%%%%%%%%%%%%%%%%%%%%%%%%%%%%%%%%%%%%%%%%%%%%%%%%%%%%%%
\subsection{Primed Dirac operators}
In this subsection, we construct the fully covariant, `primed' Dirac operators for the $\ODD$ non-singlet fermions, \textit{i.e.~}the primed fermions, $(\psip_{\brp}^{\,\alpha},\rhop^{\alpha})$.  As the analysis is parallel to the previous subsection on the unprimed fermions, we skip the details and present only the main results.\\

The primed master semi-covariant derivative is
\be
\cDp_{A}:=\partial_{A}+\Gamma_{A}+\Phip_{A}+\brPhip_{A}=\na_{A}+\Phip_{A}+\brPhip_{A}=\Dp_{A}+\Gamma_{A}\,,
\label{cDA}
\ee
where  
\be
\Dp_{A}:=\partial_{A}+\Phip_{A}+\brPhip_{A}\,,
\ee
and
\be
\ba{ll}
\Phip_{Apq}=-\brVp_{Bp}\na_{A}\brVp^{B}{}_{q}\,,~~~~&~~~~
\brPhip_{A\brp\brq}=-\Vp_{B\brp}\na_{A}\Vp^{B}{}_{\brq}\,,\\
\multicolumn{2}{c}{\Gamma_{ABC}=-\Vp_{B}{}^{\brp}\Dp_{A}\Vp_{C\brp}-\brVp_{B}{}^{p}\Dp_{A}\brVp_{Cp}\,.}
\ea
\ee
It satisfies
\be
\ba{cc}
\cDp_{A}\Vp_{B\brp}=0\,,~~~~~~~~~~~
\cDp_{A}\brVp_{Bp}=0\,,~~~&~~~~\cDp_{A}d=\na_{A}d=\partial_{A}d+\half\Gamma^{B}{}_{BA}=0\,,\\
\cDp_{A}P_{BC}=\na_{A}P_{BC}=0\,,~~~&~~~~\cDp_{A}\brP_{BC}=\na_{A}\brP_{BC}=0\,,
\ea
\label{VVdp}
\ee
and like (\ref{JeeCC}),
 \be
 \ba{llll}
\cDp_{A}\cJ_{BC}=0\,,~~~~&~~~~\cDp_{A}\eta_{pq}=0\,,~~~~&~~~~\cDp_{A}\breta_{\brp\brq}=0\,,~~~~&~~~~{}\\
\cDp_{A}C_{\alpha\beta}=0\,,~~~~&~~~~\cDp_{A}\brC_{\bralpha\brbeta}=0\,,~~~~&~~~~
\cDp_{A}(\gamma^{p})^{\alpha}{}_{\beta}=0\,,~~~~&~~~~
\cDp_{A}(\brgamma^{\brp})^{\bralpha}{}_{\brbeta}=0\,.
\ea
\ee
Further, in analogy to (\ref{SF}), we have
\be
\ba{l}
P^{AB}S_{AB}=-\brFp_{AB\brp\brq}\Vp^{A\brp}\Vp^{B\brq}- \half\Gamma_{ABC}\Gamma^{AB}{}_{D}P^{CD}\,,\\
\brP^{AB}S_{AB}=-\Fp_{ABpq}\brVp^{Ap}\brVp^{Bq}- \half\Gamma_{ABC}\Gamma^{AB}{}_{D}\brP^{CD}\,,
\ea
\label{SFp}
\ee
where the primed field strengths  are, in an  identical fashion to (\ref{FPhi}),
\be
\ba{ll}
\Fp_{AB}=\partial_{A}\Phip_{B}-\partial_{B}\Phip_{A}+\left[\Phip_{A},\Phip_{B}\right]\,,~~~~&~~~~
\brFp_{AB}=\partial_{A}\brPhip_{B}-\partial_{B}\brPhip_{A}+\left[\brPhip_{A},\brPhip_{B}\right]\,.
\ea
\ee
~\\
\indent Finally, the fully  covariant primed Dirac operators  are 
\be
\ba{llll}
\gammap^{A}\cDp_{A}\rhop\,,~~~~&~~~~\gammap^{A}\cDp_{A}\psip_{\brp}\,,~~~&~~\Vp^{A}{}_{\brp}\cDp_{A}\rhop\,,~~~&~~\Vp^{A\brp}\cDp_{A}\psip_{\brp}=\cDp_{A}\psip^{A}\,.
\ea
\label{primedDirac}
\ee
Here   we  set for simplicity,
\be
\ba{ll}
\psip_{A}:=\Vp_{A}{}^{\brp}\psip_{\brp}\,,~~~~&~~~~\gammap^{A}:=\brVp^{A}{}_{p}\gamma^{p}\,,
\ea
\ee
such  that 
\be
\ba{ll}
\Vp^{A}{}_{\brp}\psip_{A}=-\psip_{\brp}\,,~~~~&~~~~\left\{\gammap^{A},\gammap^{B}\right\}=-2\brP^{AB}\,.
\ea
\ee 
Writing explicitly we have
\be
\ba{l}
\cDp_{A}\rhop=\Dp_{A}\rhop=(\partial_{A}+\quarter\Phip_{Apq}\gamma^{pq})\rhop\,,\\
\cDp_{A}\psip_{\brp}=\Dp_{A}\psip_{\brp}=(\partial_{A}+\quarter\Phip_{Apq}\gamma^{pq})\psip_{\brp}
+\brPhip_{A\brp}{}^{\brq}\psip_{\brq}\,,\\
\cDp_{A}\psip_{B}=(\partial_{A}+\quarter\Phip_{Apq}\gamma^{pq})\psip_{B}+\Gamma_{AB}{}^{C}\psip_{C}\,.
\ea
\ee
%~\\

%%%%%%%%%%%%%%%%%%%%%%%%%%%%%%%%%%%%%%%%%%%%%%%%%%%%%%%%%%%%%%%%%%%%%%%%%%%%%%%%%%%%%%%%%%%%%%%%%%%%%%%%%%%%%%%%%%%%%%%%%%%%%%%%%%%%%%%%%%%%%%%%%%%%%%%%%%%%%%%%%%%%%%%%%%%%%%%%%%%%%%%%%%%%%%%%%%%%%%%%%%%%%%%%%%%%%%%%%%%%%%%%%%%%%%%%%%%%%%%%%%%%%%%%%%%%%%%%%%%%%%%%%%%%%%%%%%
\subsection{Chirality change under $\ODD$ T-duality}
In order to discuss the  $\ODD$ transformations of the  primed fermions, we first recall  
the $\ODD$ transformation rule of the primed double-vielbein (\ref{avecp}),
\be
\ba{ll}
\Vp_{A\brp}~~\longrightarrow~~M_{A}{}^{B}\brL_{\brp}{}^{\brq}\Vp_{B\brq}\,,~~~~&~~~~\brVp_{Ap}~~\longrightarrow~~M_{A}{}^{B}L_{p}{}^{q}\brVp_{Bq}\,,
\ea
\label{avecpr}
\ee
where, from (\ref{Mpr}) and (\ref{LbrL}), the $\ODD$ group element and the associated local Lorentz transformations are given by 
\be
\ba{ll}
M_{A}{}^{B}=\left(\ba{cc}\mba^{\mu}{}_{\nu}&\mbb^{\mu\sigma}\\ 
\mbc_{\rho\nu}&\mbd_{\rho}{}^{\sigma}\ea\right)\,,~~~&~~~
{\ba{l}
L=e^{-1}\left[\mba^{t}+(g-B)\mbb^{t}\right]\left[\mba^{t}-(g+B)\mbb^{t}\right]^{-1}e\,,\\
\brL=(\bre^{-1}e)L^{-1}(e^{-1}\bre)\,.\ea}
\ea
\ee
We also recall  the covariance of the gamma matrices (\ref{gLS}),
\be
\gamma^{q}S_{L}L_{q}{}^{p}=S_{L}\gamma^{p}\,.
\ee
It is then clear that, with the full covariance of the primed Dirac operators (\ref{primedDirac}),  the primed fermions transform under  $\ODD$ T-duality as follows,
\be
\ba{lll}
\left(\ba{c}
\rhop\\
\psip_{\brp}\\
\gammap^{A}\cDp_{A}\rhop\\
\gammap^{A}\cDp_{A}\psip_{\brp}\\
\Vp^{A}{}_{\brp}\cDp_{A}\rhop\\
\cDp_{A}\psip^{A}
\ea
\right)~~~&~~~\longrightarrow~~~&~~~
\left(
\ba{c}
S_{L}\rhop\\
\brL_{\brp}{}^{\brq}S_{L}\psip_{\brq}\,,\\
S_{L}\gammap^{A}\cDp_{A}\rhop\\
\brL_{\brp}{}^{\brq}S_{L}\gammap^{A}\cDp_{A}\psip_{\brq}\\
\brL_{\brp}{}^{\brq}\Vp^{A}{}_{\brq}S_{L}\cDp_{A}\rhop\\
S_{L}\cDp_{A}\psip^{A}
\ea
\right)\,.
\ea
\ee
Thus, from (\ref{brS5}) and (\ref{detL}),
\be
\ba{ll}
\gamma^{(D+1)}S_{L}=\det(L)\,S_{L}\gamma^{(D+1)}\,,~~~~&~~~~
\det(L)=\dis{\frac{\det\left[\mba+\mbb(g+B)\right]}{\det\left[\mba-\mbb(g-B)\right]}}=\pm 1\,,
\ea
\label{FLIPcon}
\ee
when $\det(L)=-1$,  the primed fermions flip their chiralities.  Otherwise not. \\

For example, on a flat background  ($g=\eta$, $B=0$), we may set both $\mba$ and $\mbb g$ to be diagonal with  the eigenvalues, zero or one  only, in an exclusive manner such that  $\mba+\mbb g=1$. This choice corresponds to the usual discrete T-duality along toroidal directions.  In this case, we get  $\det(L)=(-1)^{\sharp_{\mba}}$ where $\sharp_{\mba}$ counts the number of zero  eigenvalues in the matrix, $\mba$,  and hence the number of toroidal directions on which T-duality is performed.  Thus, our formula is consistent with the well-known knowledge   that performing odd number of T-duality on flat backgrounds  exchanges type IIA and IIB superstrings.  \\
\newpage

%%%%%%%%%%%%%%%%%%%%%%%%%%%%%%%%%%%%%%%%%%%%%%%%%%%%%%%%%%%%%%%%%%%%%%%%%%%%%%%%%%%%%%%%%%%%%%%%
%%%%%%%%%%%%%%%%%%%%%%%%%%%%%%%%%%%%%%%%%%%%%%%%%%%%%%%%%%%%%%%%%%%%%%%%%%%%%%%%%%%%%%%%%%%%%%%%
\subsection{Reduction to $D$ dimension}
Upon the level matching constraint (\ref{constraint}), with the explicit forms of the  double-vielbeins  (\ref{Vform}), (\ref{Vpform}),  the covariant DFT Dirac operators (\ref{unprimedDirac}), (\ref{primedDirac}) reduce to more familiar $D$-dimensional  expressions  within the  Riemannian setup, 
\be
\ba{l}
\sqrt{2} \gamma^{A}\cD_{A}\rho\equiv \gamma^{m} \left( \partial_{m} \rho + \frac{1}{4} \omega_{m n p} \gamma^{n p} \rho + \frac{1}{24} H_{m n p} \gamma^{n p} \rho - \partial_{m} \phi \rho \right) \,,
\\
\sqrt{2}\gamma^{A}\cD_{A}\psi_{\brp}\equiv \gamma^{m} \left( \partial_{m} \psi_{\brp} + \frac{1}{4} \omega_{m n p} \gamma^{n p} \psi_{\brp} + \bar{\omega}_{m \brp\brq} \psi^{\bar{q}} + \frac{1}{24} H_{m n p} \gamma^{n p} \psi_{\brp} + \half H_{m \brp \brq} \psi^{\brq} - \partial_{m} \phi \psi_{\brp}
\right)\,,
\\
\sqrt{2}\brV^{A}{}_{\brp}\cD_{A}\rho\equiv \partial_{\brp} \rho + \frac{1}{4} \omega_{\brp q r} \gamma^{q r} \rho + \frac{1}{8} H_{\brp q r} \gamma^{q r} \rho \,,
\\
\sqrt{2}\cD_{A}\psi^{A}\equiv \partial^{\brp} \psi_{\brp} + \frac{1}{4} \omega_{\brp qr} \gamma^{q r} \psi^{\brp} +\bar{\omega}^{\brp}{}_{\brp \brq} \psi^{\brq} +  \frac{1}{8} H_{\brp q r} \gamma^{q r} \psi^{\brp} - 2 \partial_{\brp} \phi \psi^{\brp} \,,
\\
\sqrt{2}\gammap^{A}\cDp_{A}\rhop\equiv \gamma^{m} \left( \partial_{m} \rhop + \frac{1}{4} \omega_{mnp} \gamma^{np} \rhop - \frac{1}{24} H_{mnp} \gamma^{np} \rhop - \partial_{m} \phi \rhop\right) \,,
\\
\sqrt{2}\gammap^{A}\cDp_{A}\psip_{\brp}\equiv \gamma^{m} \left( \partial_{m}\psip_{\brp} + \frac{1}{4} \omega_{mnp} \gamma^{np} \psip_{\brp} + \bar{\omega}_{m \brp}{}^{\brq} \psip_{\brq} - \frac{1}{24}H_{mnp} \gamma^{np}
\psip_{\brp} - \half H_{m \brp \brq} \psip^{\brq} -\partial_{m}\phi \psip_{\brp}
\right)\,,
\\
\sqrt{2}\Vp^{A}{}_{\brp}\cDp_{A}\rhop\equiv \partial_{\brp} \rhop + \frac{1}{4} \omega_{\brp q r} \gamma^{q r} \rhop - \frac{1}{8} H_{\brp q r} \gamma^{q r} \rhop \,,
\\
\sqrt{2}\cDp_{A}\psip^{A}\equiv\partial^{\brp} \psip_{\brp} + \frac{1}{4} \omega^{\brp}{}_{qr} \gamma^{qr} \psip_{\brp} + \bar{\omega}^{\brp}{}_{\brp\brq} \psip^{\brq} - \frac{1}{8} H_{\brp qr} \gamma^{qr} \psip^{\brp} -2 \partial_{\brp} \phi \psip^{\brp}\,,
\ea
\ee
where, with the $D$-dimensional standard diffeomorphism covariant derivative, $\trd_{\mu}$,  we set  $\partial_{p}=(e^{-1})_{p}{}^{\mu}\partial_{\mu}$,  $\,\partial_{\brp}=(\bre^{-1})_{\brp}{}^{\mu}\partial_{\mu}$, 
$\,\omega_{\mu pq}=(e^{-1})_{p}{}^{\nu}\trd_{\mu}e_{\nu q}$, $\,\bar{\omega}_{\mu \brp\brq}=(\bre^{-1})_{\brp}{}^{\nu}\trd_{\mu}\bre_{\nu \brq}$, \textit{\,etc.}   \\

In fact, the above expressions are precisely what appear  in type IIA and IIB supergravities~\cite{Coimbra:2011nw}, where $\psi_{\brp}$ and $\psip_{\brp}$ are gravitinos in string frame, while $\rho$ and $\rhop$ are `DFT-dilatinos' corresponding  to the  superpartner of the  DFT-dilaton, $d=\phi-\half\ln\sqrt{-g}$. \\\newpage

%%%
%%%%%%%%%%%%%%%%%%%%%%%%%%%%%%%%%%%%%%%%%%%%%%%%%%%%%%%%%%%%%%%%%%%%%
%%%%%%%%%%%%%%%%%%%%%%%%%%%%%%%%%%%%%%%%%%%%%%%%%%%%%%%%%%%%%%%%%%%%%
%%%%%%%%%%%%%%%%%%%%%%%%%%%%%%%%%%%%%%%%%%%%%%%%%%%%%%%%%%%%%%%%%%%%%
\section{Summary and comments\label{SECconclusion}}   
In summary, based on the stringy differential geometry that is  characterized by the semi-covariant derivative~\cite{Jeon:2011cn},  we have incorporated  fermions, like gravitino and dilatino,  into double field theory in a manifestly covariant manner with regard  to  all the symmetries in Table \ref{TABsymmetry}, \textit{i.e.}   $\ODD$ T-duality,  double-gauge symmetry and a pair of local Lorentz symmetries.  We have shown    that  in general there are two types  of fermions in double field theory: $\ODD$  singlet   and non-singlet (unprimed and primed). For each type,  we have  constructed relevant  covariant Dirac operators, (\ref{unprimedDirac}) and (\ref{primedDirac}).  Especially,   we have derived a  necessary and sufficient condition for the primed fermions to flip their chiralities under  $\ODD$ T-duality (\ref{FLIPcon}), that   depends on both the $\ODD$ group element  and the background fields.  \\

In this paper, we have chosen the primed fermions, $(\rhop^{\alpha},\psip_{\brp}^{\,\alpha})$, to carry the same local Lorentz indices as the unprimed fermions, $(\rho^{\alpha},\psi_{\brp}^{\,\alpha})$. The alternative choice is also possible:  If we let the primed fermions  have the opposite local Lorentz structure  like $(\rhop^{\bralpha},\psip_{p}^{\,\bralpha})$, their fully covariant Dirac operators are, with 
$\brgammap^{A}:=\Vp^{A}{}_{\brp}\brgamma^{\brp}$,
\be
\ba{l}
\sqrt{2}\brgammap^{A}\cDp_{A}\rhop\equiv \brgamma^{\bar{m}} \left( \partial_{\brm} \rhop + \frac{1}{4} \bromega_{\brm \brp \brq} \brgamma^{\brp\brq} \rhop - \frac{1}{24} H_{\brm \brp \brq} \brgamma^{\brp \brq} \rhop - \partial_{\brm} \phi \rhop\right) \,,
\\
\sqrt{2}\brgammap^{A}\cDp_{A}\psip_{p}\equiv \brgamma^{\brm} \left( \partial_{\brm}\psip_{p} + \frac{1}{4} \bar{\omega}_{\brm \brp \brq} \brgamma^{\brp \brq} \psip_{p} + \omega_{\brm p}{}^{q} \psip_{q} -\frac{1}{24}H_{\brm \brp \brq} \brgamma^{\brp \brq} \psip_{p} - \half H_{\brm p q} \psip^{q} -\partial_{\brm}\phi \psip_{p}
\right)\,,
\\
\sqrt{2}\brVp^{A}{}_{p}\cDp_{A}\rhop\equiv\partial_{p} \rhop + \frac{1}{4} \bar{\omega}_{p \brp \brq} \brgamma^{\brp \brq} \rhop - \frac{1}{8} H_{p \brp \brq} \brgamma^{\brp \brq} \rhop\,,
\\
\sqrt{2}\brVp^{Ap}\Dp_{A}\psip_{p}\equiv \partial^{p} \psip_{p} + \frac{1}{4} \bar{\omega}^{p}{}_{\brp \brq} \brgamma^{\brp \brq} \psip_{p}  + \omega^{p}{}_{pq} \psip^{q}- \frac{1}{8} H_{p\brp \brq} \brgamma^{\brp \brq} \psip^{p} -2 \partial_{p} \phi \psip^{p}\,.
\ea
\ee
%However, these seem irrelevant to type IIA and IIB supergravity  (\textit{e.g.} see Eq.(2.8) in \cite{Coimbra:2011nw}).   \\
~\\
\indent It is worth while to note that, the distinction between the primed and unprimed double-vielbein is arbitrary: if we set one to be $\ODD$ vector, like (\ref{vecunp}), then the other is not a vector anymore, like (\ref{avecp}).  Thus, $\ODD$ may  act on the unprimed fermions nontrivially while leaving the primed fermions singlet.   \\

So far, we have focused on the gravitational interpretation of the unprimed and primed fermions. However, we may also regard $\rho$ or $\rhop$ as gaugino  and couple them  to the Yang-Mills double field theory~\cite{Jeon:2011kp}.\\

Up to the RR sector (for related works see \textit{e.g.~}\cite{Fukuma:1999jt,Hohm:2011dv,Coimbra:2011nw}), the unifying supersymmetric double field theory reformulation of type IIA and IIB supergravities will, when constructed~\cite{JLPsDFT},  contain  the following  leading order terms (see also \cite{Kleinschmidt:2004dy}), 
\be
e^{-2d}\left(P^{AB}S_{AB}+\brrho\gamma^{A}\cD_{A}\rho+2\brpsi^{A}\cD_{A}\rho+\brpsi^{A}\gamma^{B}\cD_{B}\psi_{A}+ \brrhop\gammap^{A}\cDp_{A}\rhop+2\brpsip^{A}\cDp_{A}\rhop+\brpsip^{A}\gammap^{B}\cDp_{B}\psip_{A}\right)\,.
\ee
In particular,  the  complete supersymmetric double field theory  will manifest not only  $\ODD$ T-duality and double-gauge symmetry,  but also a pair of local Lorentz symmetries.  It will be of interest to identify  the pair of  local Lorentz symmetries directly from the string worldsheet or $\cM$-theory points of view~\cite{Hull:2004in,Hull:2006va,Berman:2007xn,Berman:2007yf,Berman:2010is,Berman:2011pe,Copland:2011yh,Thompson:2011uw,Albertsson:2011ux,Kan:2011vg,Aldazabal:2011nj}. 
%%%%%%%%%%%%%%%%%%%%%%%%%%%%%%%%%%%%%%%%%%%%%%%%%%%%%%%%%%%
%%%%%%%%%%%%%%%%%%%%%%%%%%%%%%%%%%%%%%%%%%%%%%%%%%%%%%%%%%%%%
\section*{Acknowledgements} We wish to thank  David Berman and  Daniel Waldram  for discussion.    The work was supported by the National Research Foundation of Korea\,(NRF) grants  funded by the Korea government\,(MEST) with the Grant No.  2005-0049409 (CQUeST)  and No.  2010-0002980.  The work by IJ is partially supported by NRF though the Korea-CERN theory collaboration.

\end{document}